\documentclass[preprint]{aastex}\usepackage{natbib}
\slugcomment{Version 10 Aug 04}
\shorttitle{Permanent All Sky Survey}
\shortauthors{Deeg et al.}
\begin{document}
\title{An All-Sky Survey for the Detection of Transiting Extrasolar 
Planets and for Permanent Variable Star Tracking}
\author{Hans J. Deeg, Roi Alonso, and Juan A. Belmonte}
\affil{Instituto de Astrof\'{\i}sica de Canarias, E-38205 La Laguna, Tenerife, 
Spain}
\email{hdeeg@iac.es}
\author{Khalid Alsubai, Keith Horne}
\affil{University of St Andrews, Fife KY16 9SS, Scotland, UK}
\author{Laurance R. Doyle}
\affil{SETI Institute, 515 N. Wishman Ave., Mountain View, CA 94043, USA}
\begin{abstract}
An overview is given of the Permanent All Sky Survey (PASS) project. The
primary goal of PASS is the detection of all transiting giant planets in
the entire sky, complete for stellar systems of magnitudes $\approx$ 5.5 -- 10.5. Since the sample stars are fairly
bright and relatively close, planets detected by PASS would be ideally suited
for any follow-up study with ground- or space-based instrumentation.  The
survey would also allow the pursuit of a variety of work on temporal
 astronomical phenomena of any kind, and is intended to lead to a permanent 
 all-sky tracking of variable stars with high temporal resolution. 
 The instrument consists of arrays of CCD cameras with wide-field 
 optics that cover the entire sky visible from their observing 
 locations. Calculations of the instrument's noise
sources and subsequent simulations indicate that the proposed design is able 
to achieve the
prime objective of a full-sky survey for transits. An equation for the 
signal-to-noise ratio from photometry of unguided stellar 
images is given in the appendix, together with equations for the detection 
probability of planetary transits  based on the observational coverage and 
the intrument's duty cycle.
\end{abstract}

\section{Introduction}
Since the detection of the first transiting planet in the year 1999  
\citep{charb+00,henry00}, a wide variety
of experiments to detect such planets have been proposed, or are 
already in an operational phase (see \citealt{horne03} for an overview\footnote{http://star-www.st-and.ac.uk/~kdh1/transits/table.html gives an actualized list with links to the experiments}). 
The interest in these detections is motivated by the relatively large
 amount of
 information to be gained from transiting planets and by the
 possibilities these planets provide for gaining further insights from
 follow-up studies. These allow significantly more possibilities than 
 for the majority of planets, which have been detected through radial 
 velocity measurements\footnote{See http://www.obspm.fr/encycl/encycl.html for an actual list of extrasolar planets and the method of their discovery}. The first known transiting planet, around HD\,209458,
 is currently by far the best studied extrasolar planet. 
 Together with
 results from its detection by radial velocities, all its major planetary
 and orbital parameters have been determined. Also, spectroscopic
 observations during  transits of the planet have allowed the first 
 detection of the components of an extrasolar planet's atmosphere \citep{cbn+02}, 
 and a host of further studies have been proposed (e.g., \citealt{char02}, 
 \citealt{seagroves+03}). The next
 three transiting planets to be discovered, based on data from OGLE survey \citep{upz+02,uzs+02,ups+03}, on the other hand 
  orbit much fainter
  ($I=$14.4 to 15.7 mag) stars, and observations for their independent detection through radial
 velocities, or further characterizing studies have been significantly more difficult \citep{ktj+03,kts+03,kts+04,bps+04}.
  Finally, for
  planets detected  from radial velocities alone, only their orbital parameters and 
  lower limits of
  their masses are known.
  The major factor that allows the wide variety of studies on HD\,209458b
  is the brightness of the central star $(V=7.65$ mag), which allows, for
  example, the acquisition of spectra with very high signal-to-noise (S/N) during 
  transits. High S/N is required here in order to extract the very small 
  differences
  between spectra taken on-  or off-transit.  In general, all measurements and 
  diagnostics whose S/N is dominated by  the source's photon-noise will profit 
  from observing the brightest sources possible. An important goal, therefore,
  is to detect further transiting systems around relatively bright stars.  
  As the number of bright stars in the sky is limited, the optimum sample is 
  a survey of the entire sky. The major objectives of the project 
  introduced here are the detection of  {\itshape all} transiting giant planets 
  around bright stars
  with periodicities up to several weeks and the creation therefrom of
   a catalog of
  those extrasolar planets most suited for follow-up studies. 
  
  Surveying of the entire sky is also the major difference between this project 
  and nearly all other planetary transit search projects \citep{horne03}, which
  tend to observe selected celestial zones, ranging from arcminute-sized 
  zones or single stars to fields of
  several hundred square degrees. Planets detected by these experiments 
  would typically circle much fainter central stars, which offers much 
  more limited possibilities for further studies.  Furthermore, the intention 
  of this project is to continue in operation indefinitely. 
  The underlying reason for this is the establishment of a permanent photometric 
  all-sky tracking of variable stars as a service to the astronomical 
  community at large while at the same time achieving high detection 
  probabilities of long-period transiting planets.

A preliminary description of this project, denominated PASS 
(Permanent
All Sky Survey) has been given by \citet{deeg02}. Somewhat later, \citet{pepper+03} 
presented a scaling relation for all-sky surveys that
 relates survey depth with aperture size, for which they give an optimum 
 of 5 cm. PASS would not be the first experiment to perform 
 permanent all-sky surveillance, however. Several all-sky
 cameras, such as the Cerro Tololo All Sky 
 Camera (TASCA\footnote{http://www.ctio.noao.edu/$\sim$david/tasca.htm}), 
 and the CONCAM
 network\footnote{http://www.concam.net/}, have been in operation in recent 
 years. In opposite to PASS, these are based on
 single CCD cameras with fish-eye lenses, with the principal aim of
 performing  temporal surveys of meteorological conditions, such as
 cloud cover, extinction, and sky brightness;  the detection of
 meteorites, and the detection of optical counterparts to gamma-ray
 bursts (RAPTOR; \citealt{vestrand02}).
In addition, there is the All Sky
Automated Survey (ASAS, \citealt{pojma01}), with the goal of  monitoring all
stars brighter than  magnitude 14 for variability, but whose temporal
coverage is not well suited to obtaining high detection probabilities of
transiting planets, and which  is insensitive to stellar variability 
on timescales of less than one day.  The design of PASS, on the other hand, 
is driven by the acquisition of stellar light curves with sufficient 
precision and sufficient temporal resolution for transit detection. The KELT 
(Kilodegree Extremely Little Telescope, \citealt{pepper+04}) has also been 
developed for an all-sky transit survey. It uses a rather different observing 
scheme, however, based on a single wide-angle camera that achieves all-sky 
coverage by cycling through about ten different pointings. 
It should be noted there has been already one transit detection experiment,
the 
South Pole Exoplanet Transit Search \citep{caldwell03}, 
that used an instrumental set-up similar to that of PASS.    The goal of 
that project is the detection of transits in a limited
zone near the  South Celestial Pole, by employing a fixed CCD camera with
a telephoto lens operating during an entire Antarctic winter
season. 

This paper is intended to be the first of several publications about PASS,
covering both the instrumental design and, later, the scientific
results.
 In this paper, the foundations for the objectives of PASS are
 given in the next section. The instrumental and operational
 set-up of PASS (Section 3)  and its expected performance are then introduced
 (Sections 4--5). Simulations to evaluate the performance of the experiment
 are described in Section 6, and our conclusions are given (Section 7) indicating the
 next steps in the deployment of PASS.
 
 \section{Objectives of PASS}
 The principal goal of PASS is the detection of all giant planets
 transiting bright stars between magnitudes of about 5.5 -- 10.5, with transits
 deeper than $\approx$ 10 mmag.  The initial objective is the detection
 of all of those planets with periods of up to one week. A
 considerable detection probability is also being obtained for planets with periods of up to several weeks. 
 This probability depends principally on the observational coverage, about which
 more details are given in Section 4. Since observations from at
 least two sites are needed for true all-sky coverage and an initial
 experiment may start on a single site only, some ranges of declination may
 remain uncovered.  The requirement of a limiting sample magnitude of about
   10.5 mag is based on the number of stars  needing to be surveyed
 for transits in order to achieve a representative quantity  (of the order of several
 tens) of exoplanet
 detections  useful for further study. 
 For operations lasting several years, and possibly of indefinite duration, 
 the detection of cooler planets with periods longer than a few weeks will also be possible. 
  
 Beyond planet detection, data about any photometric variability within 
 the instruments sensitivity will be obtained, allowing a wide range of studies, 
 potentially in these fields:
 \begin{itemize}
 \item variable stars of any kind
 \begin{itemize}
 \item the detection and follow-up of stellar variabilities 
 with low amplitudes (as low as 0.1\%, depending on stellar brightness and 
 frequency)
 \item flares
 \item the detection of supernovae
 \end{itemize}
 \item the detection of meteorites (their frequency, brightness and direction)
 \item the detection of optical counterparts to gamma-ray bursts and 
 ``optical flashes''
 of unknown origin
 \item the detection of stellar occultations by dark objects (e.g., Kuiper belt 
 objects)
 \item the discovery and follow-up of asteroids and comets
 
 \item sky quality and meteorological statistics:
 \begin{itemize}
 \item the recording of sky brightness and extinction in all directions
 \item the percentage of clear sky, clouds
 \item the recording of the directions of satellites and airplanes (e.g., intrusions 
 into protected sky area over observatory sites)
 \end{itemize}\end{itemize}
 
 Although the instrument design has not been optimized for these studies,  and there 
 are specialized experiments pursuing most of these objectives, the all-sky and 
 long-duration observing mode of PASS may be expected to produce valuable 
 contributions to many of these fields. It should be noted that for 
 most variable-star studies photometric requirements are less stringent 
 than for transit detection, and that   fainter magnitude limits can therefore be 
 achieved.

\section{The design of PASS} 
Among the wide variety of ground-based transit detection projects that have 
sprung up in recent years, the use
of large-format CCD detectors for the 
simultaneous acquisition of many stellar light curves is a common feature. Also, 
in order to obtain the required observing time, most of them operate on 
dedicated telescopes, which consequently tend to be relatively small ones. 
This is not contrary to
the requirement of high photometric precision 
if the targets are
adequately bright stars.  In the context of  current ground-based 
transit detection experiments with telescope sizes ranging from a few 
cm to 4 m, PASS is on the extreme ends in two senses 
\citep{horne03}.  First,
the field to be observed by PASS is the largest: in order to obtain  
maximum temporal coverage for the completion of an all-sky survey, 
the entire
visible sky at a given observing site (about 10\ 000 deg$^2$) is being
surveyed simultaneously. Second, PASS will employ the smallest
telescopes by adapting commercially available lenses for photographic 
SLR cameras.

\begin{figure*}[ht]  
 \begin{center} 
 \includegraphics[width=15cm]{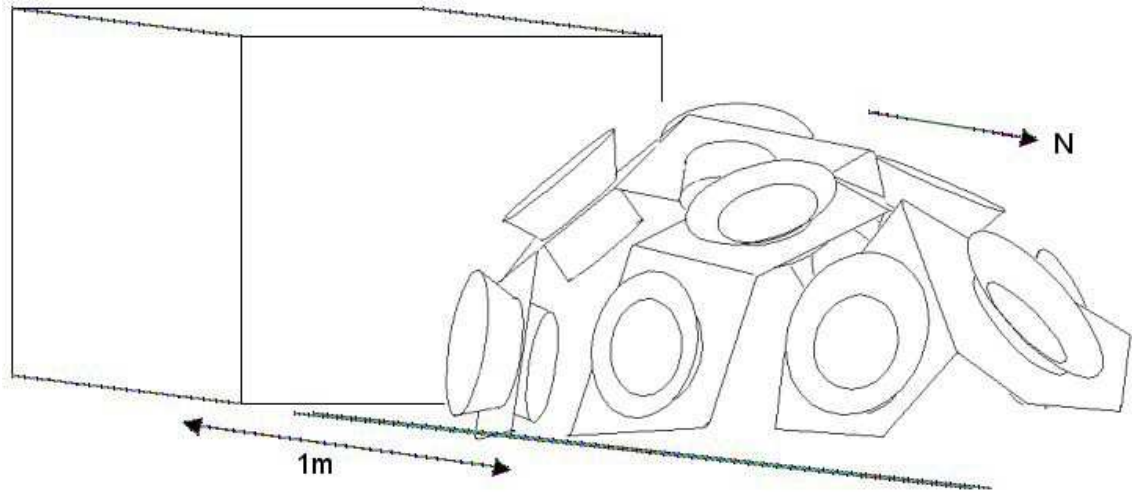} 
  \end{center}
  \caption{Schematic view of the PASS experiment, here drawn with ten cameras.
   The box in the background is the removable enclosure. An approximate 
   size scale is indicated. }
   \label{fig:design}
   \end{figure*}

\begin{figure*}[ht]  
   \begin{center}  
    \includegraphics[width=15cm]{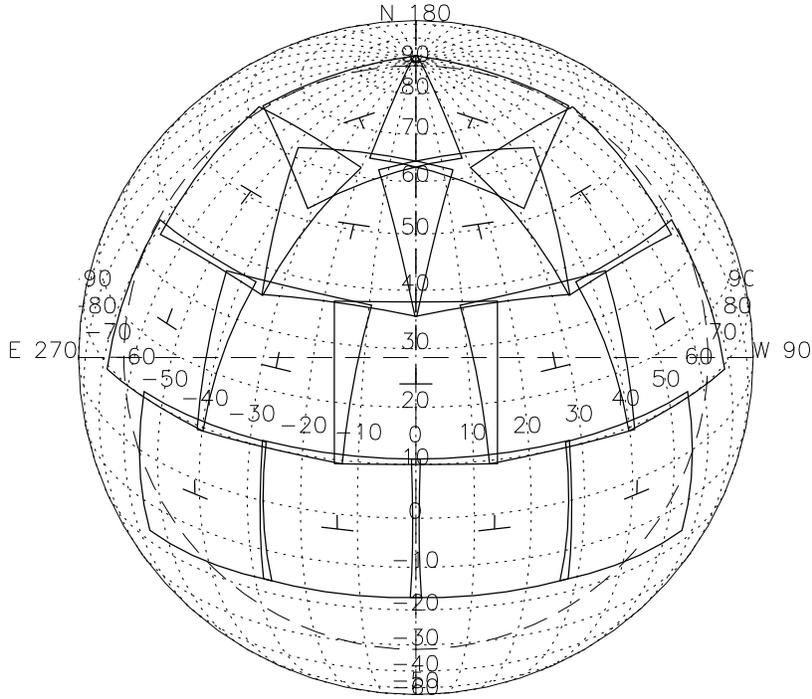} 
    \end{center}
    \caption{Local all-sky view from a location at 28.5$\degr$ N, 
    showing camera
    positions (squares) for a system of 15 units (each with a field of
    view of 28$\degr$ $\times$ 28$\degr$), in orthogonal projection. Coordinate lines are
    declination and hour angle; also indicated is an altitude of 30$\degr$ (long
    dashes around circumference). In this set-up, there is no coverage
    below declinations of $-$17.5$\degr$, as good temporal coverage of stars
    further south cannot be obtained (see Fig.~\ref{fig:cob}a).
     Further north, the sky
     is completely covered for altitudes $> 34\degr$ with an average limit
     around 30$\degr$. The altitude limit is slightly lowered in the extreme
     north, to include the  North Celestial Pole. Other camera positionings
     have been evaluated, but for a view of 28$\degr$ this is the most efficient
     one.}
     \label{fig:skycov}
     \end{figure*}

Driven by the requirement to obtain an all-sky transit survey on a 
reasonable operational timescale, a single
PASS instrument would consist of an array of 15 CCD cameras of short
focal length. With slightly overlapping fields  the cameras would
image the entire sky visible from an observing location. The
CCD cameras would be unguided (Figs~\ref{fig:design} 
and
\ref{fig:skycov}), thus ensuring  excellent mechanical and photometric 
stability of
the instrument. The instrument's baseline design
uses conventional read-out of the CCD chips, and stars would 
appear as trails in the images. Alternatively, the celestial motion
parallel to a CCD's columns may be compensated for by  synchronous
line-by-line read-out of the CCD.  The current baseline consists of 15
cameras with lenses of $f=50$ mm, as used for common high quality
SLR cameras (for 36mm film), with a CCD of appropriate size, of about 25 $\times$ 25 mm, which gives
a field of view of about $28\degr$ $\times$ $28\degr$. Figure~\ref{fig:skycov} shows 
that 15 such cameras would
give complete coverage of the sky above an altitude of around 30$\degr$. The
experiment would need to be mounted on a sturdy platform and be
covered by a completely removable enclosure. An important feature 
of the instrument will be the synchronization of the timing of the exposures 
with sidereal time (ST), such that images would always be taken  at the same 
ST and, consequently, at the same set of hour angles. Hence, on different 
nights, but at the same ST, stars would trail over exactly the same CCD pixels, 
and images taken on one night would be directly comparable to images
taken on other 
nights.  This would allow relative photometry not only among groups of stars, 
but also among data sets from many nights. Systematic errors that do \emph{not} 
vary from one night to another, such as flat-fielding errors, would therefore 
cancel out. 

The amount of data
produced would depend mainly on the size of the CCD chips (1k $\times$ 1k and 2k 
$\times$
2k designs are being evaluated) and on the level of on-line processing
 being performed. If images are co-added (accounting for the
stellar motion in the co-adding) and saved only every 500 s,
about 800 images would be generated every night, each with a size of 2
Mbyte (for a 1k $\times$ 1k chip). This would result in fairly manageable
data volumes of 1.6 Gbyte per night. Precession would cause a constant
shifting of the star trails. This may be accounted for by an
occasional recalibration of the tracks, or by  mechanical adjustment, gradually
turning the entire system  around the precession axis.  

\section{Survey coverage} 
\begin{figure*}[ht]
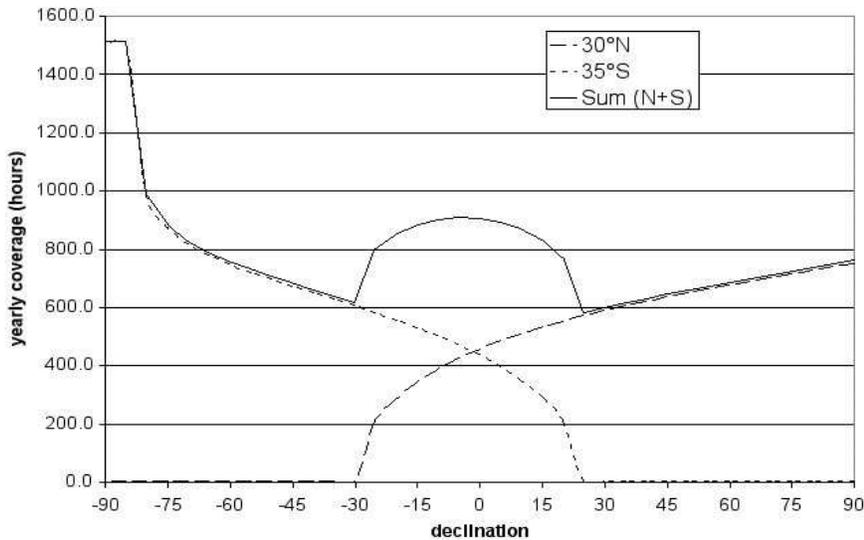
  
     \begin{center}
     a)   \includegraphics[width=13cm,height=9cm]{figcobTF.epsf}\\
     
      b)   \includegraphics[width=13cm,height=9cm]{figcobNyS.epsf}
      
      \end{center}
      \caption{a) . Yearly observational coverage of dependence on stellar
      declination, assuming a yearly total of 1500 hours of clear observing
      conditions (200 nights of 7.5 hours) for a site at 28.5$\degr$ N (Teide
      Observatory). The coverage shown here is the average for stars at any
      right ascension. Dashed line: Coverage if the entire sky above $30\degr$
      altitude is surveyed. The North Celestial Pole, at an elevation of $28.5\degr$, is not
      covered. Solid line: coverage by the 15 camera system shown in
      Figure~\ref{fig:skycov}. A small region around the  North 
      Celestial Pole is now
      circumpolar by lowering the altitude limit to $27\degr$ at very high
      northern declinations. b) As before, now showing temporal coverage
      from a northern ($30\degr$ N) and southern ($35\degr$ S) site, assuming 
      simple $30 \degr$
      altitude limits for both. If night hours do not overlap 
      among the
      sites, coverage near the celestial equator will be the sum from both
      sites, and a relatively uniform coverage (solid line) of over 600
      hr/yr is achieved over the entire sky. }
      \label{fig:cob}
      \end{figure*}

The celestial sphere above an altitude of $30\degr$ has a spatial angle of
 $1 \pi$ \emph{sterad}. Hence,  a quarter of the entire sky will
  be observed at any time. The
  amount of time that a star can be observed during the course of a year
  depends primarily on its declination and on the observatory's
  geographical latitude. Coverage also depends  strongly on the lower limit 
  for the altitude: For a $30\degr$ altitude limit at a location at $30\degr N$,
   the North Celestial Pole is permanently visible; stars at the stellar equator would
   be visible about 1/3 of the  night annually, and coverage declines
   rapidly towards southern declinations (Fig.~\ref{fig:cob}a). For the 
   proposed camera configuration from a northern location ($30\degr$ N) with 
   a southern
   declination limit of $-$17.5$\degr$ (Fig.~\ref{fig:skycov}), about 65\% of the entire sky would be
   observable with a coverage of at least 400 hr/yr. The northernmost declination range might also be observed from an array that is located geographically farther north, which allows observing this part of the sky at lower airmasses. Coverage of southern
   declinations would be achieved from at least one instrument located in that
    hemisphere. This should preferably be located at a very different longitudes to avoid
    overlapping night hours. For stars near the celestial equator, the
    coverage from two observatories in antipodal positions could then be
    doubled (Fig.~\ref{fig:cob}b), and an average  coverage of
     at least 650 hr/yr could be achieved at any declination.  

For a reliable detection of planetary candidates, and in order to suppress false 
alarms from other sources (random high sigma events, or noise of instrumental, 
meteorological, or astronomical origin),  observation of at least three
transit events would be required. The probability of detecting a given number of 
transit events for a planet of given period is primarily dependent on the 
duration of the observational coverage. This calculation is given in Appendix A.
Figure~\ref{fig:detcprob} shows the detection probability for observations 
for a duration of one and three years.
In the latter case, coverages of at least 1200 hours should be achieved, 
thus allowing high (greater than 50\%) detection probabilities for
planets with orbital periods of up to 15 days.  Longer observing spans 
would further
increase the detectable orbital periods, increase the 
confidence in
existing detections, and  to some extent lower the detectable planet sizes owing
to the observation of more transits. 

A single array at a mid-northern
or mid-southern (25--40$\degr$ latitude) site could survey about 250\,000 stars to
 $ V=10.5$ mag with sufficient precision for the detection of
giant planet transits (requiring photometry  better then 0.35\% in 900 s; 
see the following section). Similarly, a true all-sky survey could access about 400\,000 stars. Following \citet{brown03}, only about 14\% of bright field-stars are suitable for ground based transits searches (These are main-sequence stars with radii of less than 1.3R$\sun$). Assuming that about 1\% of them have 
short-orbital giant planets (e.g., \citealt{udry03}),
with a transit probability of 5\%, about 20 planets may be
detected from a 1-instrument survey, and 30 planets from an all-sky survey. It
should be emphasized that the major goal of PASS is {\it not} the detection
of large numbers of planets, but the complete detection of planets
transiting bright stars within some well-established and consistent
completeness limit.

\section{Instrumental performance} 
 It should be noted that  star trail images have a different signal to noise
  (S/N) behavior than normal  images with guided apertures. Whereas the S/N 
  of guided aperture images increases with  exposure time as $t_{\rm exp}^{1/2}$,  
  the S/N of a single star trail image converges towards a 
  fixed value for long exposure times. Consequently, in order to fill longer integrations, the acquisition and averaging of multiple shorter images is advantageous. An S/N equation describing this 
  behavior is presented in Appendix B.
Figure~\ref{fig:sn} shows the contribution from various noise sources
that have been calculated for a baseline set-up of PASS. This
consists of an $f=50$ mm lens operated at aperture $f$/2.0,\footnote{Apertures 
wider than $f$/2.0 cause vignetting or very strong flat-field gradients 
in most commercial lenses; to avoid this, $f$/2.0 was chosen for the baseline.} 
with a red cut-off filter at
725 nm (to avoid atmospheric water lines),
a front-illuminated CCD (1k $\times$ 1k size, 24 micron pixels,
with a quantum efficiency\footnote{The quantum efficiencies are obtained by
folding the CCD's QE curve with the emission spectrum of a solar star or the 
night-sky; the latter taken from \cite{lptech115}.} of 0.46 against 
light from solar-type stars and 0.39 against light from sky background), 
resulting in system quantum efficiencies (which include the cut-off 
filter and an optical efficiency of 0.8)  of 0.25 and 0.13 against
 stars and sky. Exposure times are 20 s, and for the \emph{baseline}, 
 observations of a field near the celestial equator ($\delta = 0 \degr $) 
 under a moonless sky at a dark site\footnote{For cases with moon, brightness 
 is assumed for a pointing towards the zenith with the moon at a
 zenith distance of 60$deg$.} with 21.45 mag/arcsec$^2$ has been assumed 
 \citep{lptech115}. 
 Scintillation is calculated for 1.4 airmasses at an altitude of 2400 m, 
 using the equation of \cite{young74}. Figure~\ref{fig:sn} shows that the
  three major noise sources are scintillation, photon noise from stellar sources, 
  and photon noise from sky background. Other noises, all related to the CCD chip 
  (read-out, dark signal, and digitization noise) are negligible.
With stellar photon noise as the major noise source
around 10th magnitude, the instrument may be considered as optimized for
that magnitude range, with no significant improvements possible from
the suppression of other types of  noise. 

The design of PASS, taking images from a fixed telescope and at fixed sidereal 
times will minimize any systematic errors. Photometric errors in guided 
telescopes arise from residuals in the flat-fielding correction under the 
slightly moving stellar point spread function, and from errors in the centering 
of the photometric aperture. Images taken by PASS will exhibit these same errors, 
but they will be identical in images taken at the same ST. The comparison of a 
star's brightness against a group of reference stars in identical images from 
many nights will then cancel out these errors. This will leave second order errors. 
 Temperature changes may affect the focussing, slightly varying the position 
 of stellar images. One solution may be comparative photometry only within 
 subsets of images taken at identical ST \emph{and} at similar temperatures.  
 Errors from seeing variations cannot be excluded either. However, they are 
 not expected to be significant, since the resolution of the optical system and 
 the angular size of the CCD pixels is much larger than any common seeing values. 
 First order atmospheric extinction variations are cancelled out in any
  differential photometry among nearby stars.  Since PASS fields are of very 
  wide angle, the dependence of extinction on airmass needs careful monitoring, 
  however. Second order wavelength dependent extinction variations may be
   minimized from an optimized  selection of comparison stars. This implies 
   that sample stars are classified according to their color into several groups, 
   and that photometry is performed between stars of the same group. Since PASS will
    observe the same all-sky stellar sample over a long time, the creation of
     correction functions should be able to reduce most---if not all---of these 
     errors. Such corrections can be expected to improve with prolonged 
     observations, which implies the necessity to save all raw photometry  
     of PASS for better re-reductions at later dates. 

A limiting precision for transit
detection may be defined from the requirement to detect 
transits with
at least 1\% brightness variation as 3-sigma signals in 900 s
integrations. This corresponds to requiring a precision of at least
3.5 mmag over that integration time. For the baseline set-up, this leads
to a survey limit of about 10.5 mag (see dashed line in
Fig.~\ref{fig:sn}). 
This may be considered a conservative limit, as
transits last several 
hours and would produce numerous data points of
900 s integrations,
 leading to trains of 3-sigma signals that
 should readily be classifiable 
 as transit candidates.  A bright survey limit of about 5.5 mag is given by saturation of the CCD detector. The baseline
 set-up has been calculated for 
 typical good observing conditions, as
 mentioned previously. Below, the limiting magnitudes for transit 
 detection in different
 conditions are given (conditions marked * correspond to the baseline):\\

\begin{tabbing}
\hspace*{9cm}\= \kill
Moon: none*, gibbous, half moon, full moon: \>10.5*,10.2, 9.7, 9.2\\
Apertures: $f$/1.4, $f$/1.8, $f$/2.0*:  \> 11.0, 10.7, 10.5*     \\
Exposure times: 5, 10, 20*, 30, 40, 50, 60, 100 sec\>10.6, 10.6, 10.5*, 
10.5, 10.4, 10.4, 10.3, 10.0\\
Airmass: 1.1, 1.4*, 1.7, 2.0:\>10.6, 10.5*, 10.4, 10.2\\
Declination of sample field: 0$\degr$*, 45$\degr$, 60$\degr$, 75$\degr$, 
89$\degr$, 89$\degr$ at 2 airmasses\footnotemark :\\
\>10.5*, 10.6, 10.8, 11.1, 11.1, 10.6\\
Observatory altitude (m): 0, 1000, 2400*, 4000, 0 at 2 airmasses: 
\hspace{0.5cm}10.5, 10.5, 10.5*, 10.6,9.0\\
Back-illuminated CCD with higher QE of 0.66 for stars and 0.58 for sky:
\hspace{0.5cm} 10.7 \\
CCD with double the resolution (12 $\mu$m pixels): \>10.7 \\
\end{tabbing}
\footnotetext{At mid-northern or southern latitudes, stars close to 
the celestial pole are always at high airmasses.}
The last two entries show that the use of a back-illuminated---and 
significantly more expensive---CCD with higher
quantum efficiency results in an increase in limiting magnitude of
only 0.2 mag.  Also, the use of CCDs with double the spatial resolution would  
lead only to small gains, mainly from the inclusion of smaller areas of 
sky-background within stellar apertures. The signal-to-noise calculations 
therefore give  confidence that the objective of the experiment can be reached with
the baseline design,  that objective being maintained in a variety
of differing conditions. It should be noted that the above mangitude limits 
are for an expected ``typical'' planetary transit. Lower magitude limits may be 
achieved for larger ($\ga 1.2\ R_{\rm Jup}$) planets and for many studies not related 
to planet detection, as listed in Section 2.

\section{Photometry on simulated images} 
A simulator for CCD
images by PASS has been developed
to evaluate if the noise sources predicted in the previous section can be
corrected for by photometry done on realistic images. Besides the noise sources 
described
previously, the simulations also represent stellar crowding and the
variations in inter-pixel quantum efficiency expected in
front-illuminated CCDs, as described by \citet{kaval98}. 
Figure~\ref{fig:simfield}a shows a simulated field that represents a small fraction
of the field size that would be observed by one of PASS's cameras, with 
a simulated stellar density typical for galactic latitudes of 10$\degr$ 
\citep{all73,cox99}. 
Sequences of such images have been generated in the simulations, regenerating the
appropriate noises every time and optionally accounting for the
motion of the stars from field to field. 
Stellar photometry is being performed on these image sequences
 with a program {\sc tracephot},
 written in IDL. In
 its first analysis step, an aperture mask is  built up on the
 first image, based on  known stellar positions and brightnesses 
 (from the input catalog), using the following method. The
 program starts with the brightest star and assigns any pixels under
 its trail to that star's aperture. This process is continued for
 fainter and fainter stars, but stars part of whose trails are
 already assigned to brighter stars are rejected. With this algorithm,
 the maximum number of the brightest stars is sampled in the
 field, resulting in very dense final aperture masks similar to those
 shown in Figure~\ref{fig:simfield}b. The crowding encountered in 
 the $b$ = 10$\degr$ sample field and expressed as the fraction of confused 
 apertures against stellar brightness is indicated in Table~\ref{tab:crowd}. 
 There, \emph{Nstars} is the number of stars that were in the simulation's 
 input catalog. \emph{Nassign} is the number of stars for which apertures could 
 be assigned and photometry extracted, and \emph{Nconfused} is the number 
 of confused and hence irretrievable stars, with \emph{frac$_{\rm conf}$} 
 being the fraction of confused stars. Only for magnitudes fainter than 
 $\approx$ 12 are a significant fraction of stars  rejected due to 
 confusion.
In the second step, {\sc{tracephot}} applies the
aperture mask to the first image and then, with appropriate shifts, to
the following ones. Simple aperture photometry is then performed
through these masks, resulting in a time-series for each star. 
The major factor that can be expected to degrade the observed 
photometric precision against the theoretical one, especially among
the fainter stars, is caused by the crowding that many of these stars
are suffering. Although the aperture assignment algorithm ensures that
the largest fraction of light in each pixel in a given stellar
aperture comes from the star being measured, it cannot preclude a 
significant fraction of light coming from other, fainter stars, which 
will result in additional noise.
Figure~\ref{fig:magvsmerr}
shows the {rms} error of photometry of such a time-series 
from the baseline set-up,
against the known input brightness of the stars. Comparison with the
theoretical noise figures (solid line in Fig.~\ref{fig:sn}) shows that the
photometry on stellar traces is able to extract  brightnesses
for most stars  
with  noise types  intrinsic to the images. 

\section{Conclusions and outlook}
An experiment for the detection of transiting planetary systems around all bright 
stars in the entire sky is described, within magnitudes of about 5.5 -- 10.5. 
The instrument would also provide the starting point for  permanent photometric 
tracking of variable stars of any kind. 

The field of transit detection has reached a point of inflection, 
where it becomes obvious that previous estimates of the discovery rates 
of transiting planets are turning out to be too optimistic. For example, 
\citet{horne03} predicted for the current transit experiments a total of 
10 to 100 planet detections per month, and cites six monthly detections for 
PASS and two monthly detections for the STARE 
project.\footnote{http://www.hao.ucar.edu/public/research/stare/stare.html.} 
 This latter  has been in regular operation since 2001, but to date 
 only one transiting planet has been discovered \citep{abt+04}. 
 For PASS, our estimate gives a total of 30 detected hot giant planets with 
 a single array (Section 4), which might be achieved after 3--4 years 
 of continuous operation, corresponding to an average monthly detection 
 rate of about one planet. This discrepancy  probably results from several factors: 
 simplifying assumptions in the noise characteristics that governed 
 detection limits in previous predictions (in particular unaccounted errors 
 in aperture photometry on guided telescopes), overestimates of the fraction of stars that are suitable as targets for transit surveys, and underestimates of the 
 required  observational coverage. Large amounts of transit-like events 
 can be quickly  found by any detection experiment (one of the few 
 published data at that stage are the transit-candidates from OGLE, 
 \citealt{upz+02,uzs+02}). The first cut in the selection of ``good'' candidates 
 (i.e., those worth detailed follow-up) is to seek periodicity, which needs 
 the observation of at least three transit events.  Even for short-period hot 
 giants, however, this requires  at least 300--400 hours of observations to achieve 
 detection probabilities $>$ 0.5, which corresponds to observing for at 
 least 50 nights  (see Fig.~\ref{fig:detcprob} and Appendix A). Many 
 attempts with shorter coverage have indeed failed to produce any reliable 
 detections, a typical example being \citet{street+03}. We do not wish to 
 imply that these earlier predictions or observational attempts were in any
 way faulty, 
 but they do show that the field has matured through the experience gained in 
 the course of several campaigns. 
 It is also becoming clear that the detection of a transit is 
 only the first link  in a chain to produce reliable planet detections. Estimates 
 by \citet{brown03} show that a large fraction of---if not 
 most---transit-like events in giant planet searches will be produced by several 
 stellar configurations with eclipsing binaries. Based on the experience 
 from STARE, \citet{adb+04} propose a staged approach at false alarm 
 detection:  It begins with a careful review of the transit parameters 
 following the precepts by \citet{seager+03},  and is followed by 
 observations that are simple and progress towards more demanding observations. 
 These start with transit observations in several different colors, which 
 will eliminate most eclipsing binaries, and finalize in  radial velocity 
 measurements that may lead to an independent verification of a planet. The 
 brightness of the sample surveyed by PASS will greatly facilitate 
 such observations, making first verification observations possible even from 
 rather small  20--30 cm class telescopes. Radial velocity measurements should
  not cause significant difficulties, either. Still, the large sample size of 
  PASS will need a well-coordinated effort for follow-up observations, and help 
  from the amateur community might play a very useful role for some verification 
  observations.

In this first paper, the basic set-up, S/N considerations, and simulations 
are described that indicate that the baseline design of the instrument 
could indeed fulfill the main objective. Finance has been obtained for the 
placement of a prototype system. This system comprises two CCD cameras 
with  50 mm  Nikon lenses, with an initial location at Teide Observatory on 
Tenerife, where a small dome is currently being constructed.  It is described in more detail in \citet{passproto1}. The principal 
goal of the prototype is the undertaking of a feasibility study, for which 
measurements under a variety of observing conditions and pointings will be 
obtained. The prototype will also allow refinement of  observing 
strategies and deliver real data that will aid in the development and testing 
of the future reduction pipeline. Once these observations have finished, 
it is envisaged to point the two cameras to about 62$\degr$ 
declination and hour-angles of + and - 1h 20 min (+ and - 20 $\degr$ in Fig.~\ref{fig:skycov}), thereby starting the first operating elements 
of PASS.

\section{Acknowledgement} The authors are grateful to the referee, 
whose comments led to a clearer discussion in parts of this paper. Part of this project 
 is being funded by grant
 AYA-2002-04566 of the Spanish Ministerio de Educaci\'on y Ciencia. 
 \bibliography{../HJDmain}

\begin{thebibliography}{32}
\expandafter\ifx\csname natexlab\endcsname\relax\def\natexlab#1{#1}\fi

\bibitem[{Allen(1973)}]{all73}
Allen, C. 1973, Astrophysical Quantities (London: Athlone Press)

\bibitem[{{Alonso} {et~al.}(2004{\natexlab{a}}){Alonso}, {Brown}, {Torres},
  {Latham}, {Sozzett}, {Belmonte}, {Charbonneau}, {Deeg}, {Dunham},
  {Mandushev}, {O'Donovan}, \& {Stefanik}}]{abt+04}
{Alonso}, R., {Brown}, T., {Torres}, G., {Latham}, D., {Sozzett}, A.,
  {Belmonte}, J., {Charbonneau}, D., {Deeg}, H., {Dunham}, E., {Mandushev}, G.,
  {O'Donovan}, F., \& {Stefanik}, R. 2004{\natexlab{a}}, submitted to ApJL

\bibitem[{{Alonso} {et~al.}(2004{\natexlab{b}}){Alonso}, {Deeg}, {Brown}, \&
  {Belmonte}}]{adb+04}
{Alonso}, R., {Deeg}, H.~J., {Brown}, T.~M., \& {Belmonte}, J.~A.
  2004{\natexlab{b}}, submitted to Astronomische Nachrichten

\bibitem[{{Benn} \& {Ellison}(1998)}]{lptech115}
{Benn}, C.~R., \& {Ellison}, S.~L. 1998, La Palma Tech. Note 115, Tech. rep.,
  Isaac Newton Group of Telescopes, La Palma

\bibitem[{{Bouchy} {et~al.}(2004){Bouchy}, {Pont}, {Santos}, {Melo}, {Mayor},
  {Queloz}, \& {Udry}}]{bps+04}
{Bouchy}, F., {Pont}, F., {Santos}, N.~C., {Melo}, C., {Mayor}, M., {Queloz},
  D., \& {Udry}, S. 2004, \aap, 421, L13

\bibitem[{{Bronstein} \& {Semendjajew}(1979)}]{Bronstein}
{Bronstein}, I., \& {Semendjajew}, K. 1979, Taschenbuch der Mathematik (Thun,
  Frankfurt/Main: Verlag Harri Deutsch), 701

\bibitem[{{Brown}(2003)}]{brown03}
{Brown}, T.~M. 2003, \apjl, 593, L125

\bibitem[{{Caldwell} {et~al.}(2003){Caldwell}, {Witteborn}, {Showen}, {Ninkov},
  {Martin}, {Doyle}, \& {Borucki}}]{caldwell03}
{Caldwell}, D.~A., {Witteborn}, F.~C., {Showen}, R.~L., {Ninkov}, Z., {Martin},
  K.~R., {Doyle}, L.~R., \& {Borucki}, W.~J. 2003, Astronomy in Antarctica,
  25th meeting of the IAU, Special Session 2, 18 July, 2003 in Sydney,
  Australia, 2

\bibitem[{{Charbonneau}(2003)}]{char02}
{Charbonneau}, D. 2003, in ASP Conf. Ser. 294, Scientific Frontiers in Research
  on Extrasolar Planets, eds. D. Deming, S. Seager (San Francisco: ASP), 449

\bibitem[{{Charbonneau} {et~al.}(2000){Charbonneau}, {Brown}, {Latham}, \&
  {Mayor}}]{charb+00}
{Charbonneau}, D., {Brown}, T.~M., {Latham}, D.~W., \& {Mayor}, M. 2000, ApJL,
  529, L45

\bibitem[{{Charbonneau} {et~al.}(2002){Charbonneau}, {Brown}, {Noyes}, \&
  {Gilliland}}]{cbn+02}
{Charbonneau}, D., {Brown}, T.~M., {Noyes}, R.~W., \& {Gilliland}, R.~L. 2002,
  \apj, 568, 377

\bibitem[{Cox(1999)}]{cox99}
Cox, A., ed. 1999, Allen's Astrophysical Quantities (Springer, AIP press)

\bibitem[{Deeg {et~al.}(2004)Deeg, Alonso, Belmonte, Horne, Alsubai, \&
  Doyle}]{passproto1}
Deeg, H., Alonso, R., Belmonte, J., Horne, K., Alsubai, K., \& Doyle, L. 2004,
  Astronomische Nachrichten, 325, in print

\bibitem[{{Deeg}(2002)}]{deeg02}
{Deeg}, H.~J. 2002, in Proceedings of the First Eddington Workshop on Stellar
  Structure and Habitable Planet Finding, 11 - 15 June 2001, Cordoba, Spain.
  Editor: B. Battrick, Scientific editors: F. Favata, I. W. Roxburgh, D.
  Galadi. ESA SP-485, (Noordwijk: ESA Publications Division), 273

\bibitem[{{Henry} {et~al.}(2000){Henry}, {Marcy}, {Butler}, \&
  {Vogt}}]{henry00}
{Henry}, G.~W., {Marcy}, G.~W., {Butler}, R.~P., \& {Vogt}, S.~S. 2000, \apjl,
  529, L41

\bibitem[{{Horne}(2003)}]{horne03}
{Horne}, K. 2003, in ASP Conf. Ser. 294, Scientific Frontiers in Research on
  Extrasolar Planets, eds. D. Deming, S. Seager (San Francisco: ASP), 361

\bibitem[{{Kavaldjiev} \& {Ninkov}(1998)}]{kaval98}
{Kavaldjiev}, D., \& {Ninkov}, Z. 1998, Optical Engineering, 37, 948

\bibitem[{{Konacki} {et~al.}(2003{\natexlab{a}}){Konacki}, {Torres}, {Jha}, \&
  {Sasselov}}]{ktj+03}
{Konacki}, M., {Torres}, G., {Jha}, S., \& {Sasselov}, D.~D.
  2003{\natexlab{a}}, \nat, 421, 507

\bibitem[{{Konacki} {et~al.}(2003{\natexlab{b}}){Konacki}, {Torres},
  {Sasselov}, \& {Jha}}]{kts+03}
{Konacki}, M., {Torres}, G., {Sasselov}, D.~D., \& {Jha}, S.
  2003{\natexlab{b}}, \apj, 597, 1076

\bibitem[{{Konacki} {et~al.}(2004){Konacki}, {Torres}, {Sasselov}, {Pietrzy{\'
  n}ski}, {Udalski}, {Jha}, {Ruiz}, {Gieren}, \& {Minniti}}]{kts+04}
{Konacki}, M., {Torres}, G., {Sasselov}, D.~D., {Pietrzy{\' n}ski}, G.,
  {Udalski}, A., {Jha}, S., {Ruiz}, M.~T., {Gieren}, W., \& {Minniti}, D. 2004,
  \apjl, 609, L37

\bibitem[{{Pepper} {et~al.}(2003){Pepper}, {Gould}, \& {Depoy}}]{pepper+03}
{Pepper}, J., {Gould}, A., \& {Depoy}, D.~L. 2003, Acta Astronomica, 53, 213

\bibitem[{{Pepper} {et~al.}(2004){Pepper}, {Gould}, \& {DePoy}}]{pepper+04}
{Pepper}, J., {Gould}, A., \& {DePoy}, D.~L. 2004, ArXiv Astrophysics e-prints

\bibitem[{{Pojma{\' n}ski}(2001)}]{pojma01}
{Pojma{\' n}ski}, G. 2001, in ASP Conf. Ser. 246: IAU Colloq. 183: Small
  Telescope Astronomy on Global Scales, eds. W.-P. Chen, C. Lemme, B.
  Paczynski, 53

\bibitem[{{Seager} \& {Mall{\' e}n-Ornelas}(2003)}]{seager+03}
{Seager}, S., \& {Mall{\' e}n-Ornelas}, G. 2003, \apj, 585, 1038

\bibitem[{{Seagroves} {et~al.}(2003){Seagroves}, {Harker}, {Laughlin}, {Lacy},
  \& {Castellano}}]{seagroves+03}
{Seagroves}, S., {Harker}, J., {Laughlin}, G., {Lacy}, J., \& {Castellano}, T.
  2003, \pasp, 115, 1355

\bibitem[{{Street} {et~al.}(2003){Street}, {Horne}, {Lister}, {Penny},
  {Tsapras}, {Quirrenbach}, {Safizadeh}, {Mitchell}, {Cooke}, \&
  {Cameron}}]{street+03}
{Street}, R.~A., {Horne}, K., {Lister}, T.~A., {Penny}, A.~J., {Tsapras}, Y.,
  {Quirrenbach}, A., {Safizadeh}, N., {Mitchell}, D., {Cooke}, J., \&
  {Cameron}, A.~C. 2003, \mnras, 340, 1287

\bibitem[{{Udalski} {et~al.}(2002{\natexlab{a}}){Udalski}, {Paczynski},
  {Zebrun}, {Szymaski}, {Kubiak}, {Soszynski}, {Szewczyk}, {Wyrzykowski}, \&
  {Pietrzynski}}]{upz+02}
{Udalski}, A., {Paczynski}, B., {Zebrun}, K., {Szymaski}, M., {Kubiak}, M.,
  {Soszynski}, I., {Szewczyk}, O., {Wyrzykowski}, L., \& {Pietrzynski}, G.
  2002{\natexlab{a}}, Acta Astronomica, 52, 1

\bibitem[{{Udalski} {et~al.}(2003){Udalski}, {Pietrzynski}, {Szymanski},
  {Kubiak}, {Zebrun}, {Soszynski}, {Szewczyk}, \& {Wyrzykowski}}]{ups+03}
{Udalski}, A., {Pietrzynski}, G., {Szymanski}, M., {Kubiak}, M., {Zebrun}, K.,
  {Soszynski}, I., {Szewczyk}, O., \& {Wyrzykowski}, L. 2003, Acta Astronomica,
  53, 133

\bibitem[{{Udalski} {et~al.}(2002{\natexlab{b}}){Udalski}, {Zebrun},
  {Szymanski}, {Kubiak}, {Soszynski}, {Szewczyk}, {Wyrzykowski}, \&
  {Pietrzynski}}]{uzs+02}
{Udalski}, A., {Zebrun}, K., {Szymanski}, M., {Kubiak}, M., {Soszynski}, I.,
  {Szewczyk}, O., {Wyrzykowski}, L., \& {Pietrzynski}, G. 2002{\natexlab{b}},
  Acta Astronomica, 52, 115

\bibitem[{{Udry} {et~al.}(2003){Udry}, {Mayor}, \& {Queloz}}]{udry03}
{Udry}, S., {Mayor}, M., \& {Queloz}, D. 2003, in ASP Conf. Ser. 294,
  Scientific Frontiers in Research on Extrasolar Planets, eds. D. Deming, S.
  Seager (San Francisco: ASP), 17

\bibitem[{{Vestrand} {et~al.}(2002){Vestrand}, {Borozdin}, {Brumby},
  {Casperson}, {Fenimore}, {Galassi}, {McGowan}, {Perkins}, {Priedhorsky},
  {Starr}, {White}, {Wozniak}, \& {Wren}}]{vestrand02}
{Vestrand}, W.~T., {Borozdin}, K.~N., {Brumby}, S.~P., {Casperson}, D.~E.,
  {Fenimore}, E.~E., {Galassi}, M.~C., {McGowan}, K., {Perkins}, S.~J.,
  {Priedhorsky}, W.~C., {Starr}, D., {White}, R., {Wozniak}, P., \& {Wren},
  J.~A. 2002, in Proc. SPIE, 4845, 126

\bibitem[{Young(1974)}]{young74}
Young, A. 1974, in Methods of Experimental Physics, Vol. 12, Astrophysics, Part
  A, ed. N.~Carleton (New York: Academic)

\end{thebibliography}
 
 \appendix
\section{Probability of observing a required number of transits}
The following calculation gives
 the probability that at least $N_{\rm tr}$ transits of a transiting 
system are being observed, in observations spanning a time $T_{\rm obs}$, 
during which a fraction $f_{\rm cov}$ of temporal coverage  is achieved 
(thereby giving an observational coverage of $T_{\rm cov} = T_{\rm obs}
 f_{\rm cov}$). 
 For a transiting planet with period $P$, the 
 probability, $p_{\rm tr}$, of observing at least part of a transit  
 with observations lasting for \emph{one} period ($T_{\rm obs} = P$) 
 is given by
 \begin{equation}p_{\rm tr} = \frac{(\mbox{time observed within }P)}{P} = f_{\rm cov};\end{equation}
e.g., $p_{\rm tr} $ is identical to the duty cycle, $f_{\rm cov}$. 
The probability of missing the transit is $p_{\rm notr}=1-f_{\rm
cov}$. For multiple transits, the probability of observing exactly $k$
transits within a time $T_{\rm obs} = n \cdot P$ is then represented 
by the binominal distribution: 
\begin{eqnarray}p_{k {\rm{~transits~in~}} {\rm time} n\cdot P} &=& {
n\choose k} p_{\rm tr}^k p_{\rm notr}^{n-k} \nonumber\\  
 &                  =& {n \choose k}f_{\rm cov}^k (1-f_{\rm cov})^{n-k},
 \end{eqnarray}
 where ${n \choose k}$ is the number of combinations of $k$ in $n$
 without repetition. The probability to observe \emph{at least} $N_{\rm
 min}$ transits during is then given by the summation over above terms:
 \begin{equation}p_{N_{\rm tr} \ge N_{\rm min}} = \sum_{k=N_{\rm min}}^n p_k = 
 1 - \sum_{k=0}^{N_{\rm min}-1} p_k.\end{equation}
The above equations are correct so long as the number of periods within the 
observations, $n = T_{\rm obs}/{P}$, is slightly larger than $N_{\rm min}$.
For $n-N_{\rm min} \la 5$, or duty cycles $f_{\rm cov}$ approaching 1, 
the detection probabilities are dominated by aliasing effects.

A further simplification is possible for low duty cycles, $f_{\rm cov}$,
 and large $n$ 
where we can use the relation (e.g., \citealt{Bronstein})
\begin{equation}p_{k} = {n\choose k} p^k {q}^{n-k} = \frac{\lambda^k}{k!}e^{-\lambda},\end{equation}
where $q= 1-p$ and $\lambda = n p$. Replacing $n$ by $T_{\rm obs}/{P}= T_{\rm cov}{(f_{\rm cov})} P$, and $p$ by $p_{tr}=f_{cov}$
(hence, $\lambda = T_{\rm cov}/P$) gives:

\begin{equation}p_{k} = \frac{(T_{\rm cov}/P)^k }{k!}\  e^{-T_{\rm cov}/P}. 
\end{equation}.

The detection probability, $p_{N_{\rm tr} \ge N_{\rm min}}$, can then  be 
obtained by a summation similar to the previous one.  It should be noted  
that this equation is independent of $f_{cov}$
 and in practice gives good 
results so long as $ f_{\rm cov} \la 0.5$. 
\clearpage
\section{Signal-to-noise equation for trailing star images}
In the equation to be derived, we consider only the noises dominated by photon 
statistics, from the stellar source, and the sky background. Similarly for normal 
stellar images with guided apertures, the S/N is then given by
\begin{equation}S/N = N_{\rm ph,*}/\sqrt{N_{\rm ph,*}+N_{\rm ph,s}},\end{equation}
where $N_{\rm ph,*}$ and $N_{\rm ph,s}$ are the count of detected photons from 
star and sky in a given aperture. The photon count from 
a star in a time $t$ may be expressed by its flux as
\begin{equation}N_{\rm ph,*} = \phi_{*} t.\end{equation}
For the photon count  detected from sky-background, 
in trailing star images the increase of the aperture with time has 
to be taken into account. If  $\phi_{\rm s}$ denotes the sky-flux per area 
in the detector-plane, then
\begin{equation}N_{\rm ph,s} = \phi_{\rm s} t A(t), \end{equation}
where $A(t)$ is the area of the aperture, which may be given by 
$A(t)=D^2 \pi /4+ D \omega t$, where $D$ is both the focal-plane aperture 
diameter for very short exposures and the width of a trail-shaped 
aperture for longer exposures, and $\omega$ is the velocity of a 
stellar image across the detector, given by: $\omega = 2 \pi f 
(\cos \delta)/1Sd$, where $f$ is the focal length,  $\delta$ 
the star's declination, and $1Sd$ = 86164 s is the length of a sidereal day. 
Hence,
\begin{equation}N_{\rm ph,s} = \phi_{\rm s} ( D^2 \pi /4 t +  D \omega t^2).\end{equation}
Inserting this into the first equation, the S/N equation for single trailing star 
images is then obtained:

\begin{equation}S/N=  \frac{\phi_{*} t }{\sqrt{t(  \phi_{*} + \phi_{\rm s}  D^2 \pi /4)+t^2  
\phi_{\rm s} D \omega}}  \end{equation}
For long exposures the $t^2$ term dominates, and  S/N converges towards
\begin{equation}S/N_{t \to \infty} =  \frac{\phi_{*} }{\sqrt{\phi_{\rm s} D \omega}}. \end{equation}\

Long integration times are therefore better filled with multiple shorter individual exposures.  If the noise of single images is dominated by Poisson statistics, then their measurements can be averaged, and the combined S/N for $n$ images increases by $n^{1/2}$. For an integration time of $t_{\rm int}$, which is filled by $n=t_{\rm int}/(t + t_{\rm dead})$  images, where $t_{\rm dead}$ is the 'dead-time' lost between exposures, the S/N is then given by:
\begin{equation}S/N_{t_{\rm int}} = \left( {\frac{t_{\rm int}}{t + t_{\rm dead}}}\right)^{1/2}\: \frac{\phi_{*} t }{\sqrt{t(  \phi_{*} + \phi_{\rm s}  D^2 \pi /4)+t^2  
\phi_{\rm s} D \omega}}\end{equation}

Fig.~\ref{fig:sntrail} shows the behaviour of this equation for varying values of exposure time $t$ for the PASS project, with a dead-time of $t_{\rm dead} = 4$sec. For trailing star observations with a finite dead-time between exposures, there exists  an optimum exposure time for single images  that allows the filling of any long integration by sequences of shorter images. For applications such as that presented here with the PASS project, with several further noise sources beyond photon noise from stars and sky, the quantity $t_{\rm opt}$  
 is most conventiently 
 determined by calculating the S/N as shown in Fig. 7 with varying values of 
 $t$. For the PASS baseline,  assuming a CCD read-out time of 4 s, 
 an exposure time of about 13 s optimizes S/N for stars of mag 11 at equatorial regions, 
 whereas 26 seconds are optimum for  declinations of 60$\degr$. Since the 
 relation of S/N to $t$ has a very broad peak, a uniform exposure time of 20 
 s has been assumed for the baseline, with losses of not more than 0.1 mag in limiting magnitudes at any declination.

 \clearpage
   \clearpage
     \clearpage
      \clearpage
      \begin{figure*}[ht]  
      \begin{center}%
        \includegraphics[width=15cm]{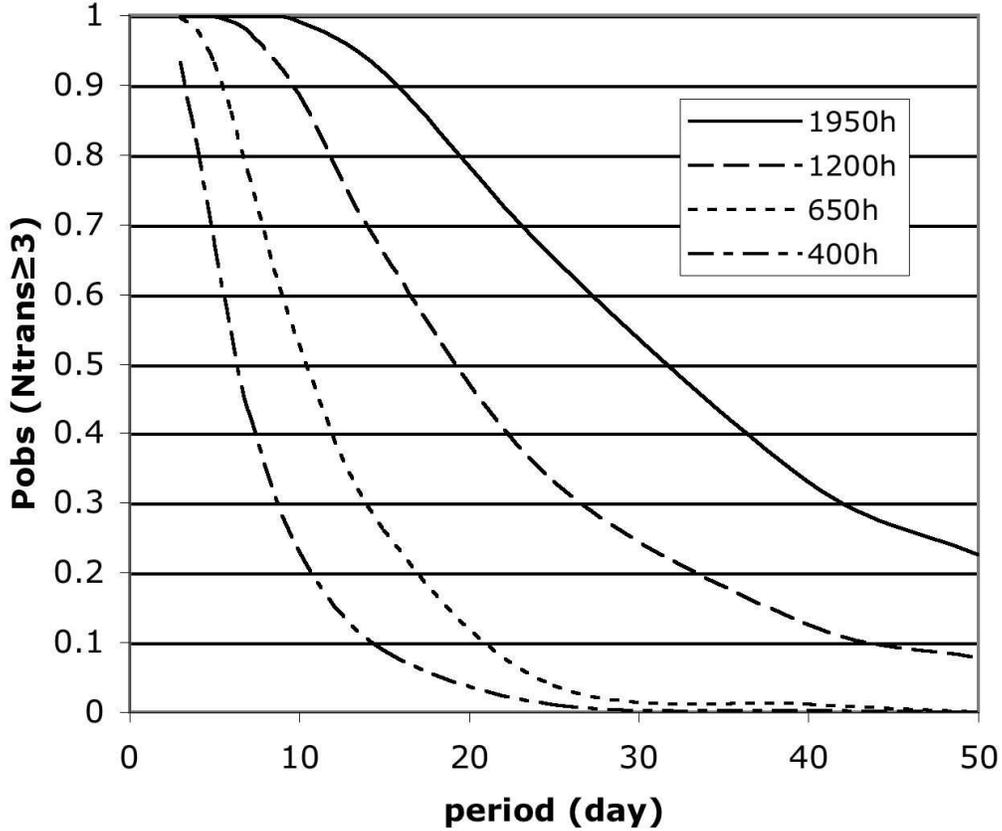} 
	 \end{center}
	 \caption{Probability, P$_{\rm obs}$, that transiting planets will be 
	 detected, depending on  their orbital period and observational coverage. The detection probability is based on the
	 requirement to observe at least three transits.  An observational coverage of 400 hours  would be achieved from a single array 
	 at a mid-northern or mid-southern site (at about 30$\degr$ N or S) during one year of observations (compare to Fig.~\ref{fig:cob}), and 1200 hours during 3 years. 650 hours of observational coverage can be achieved in one year for stars at any declination from combined  observations with a northern and southern array \emph{without}  overlapping night hours; 1950 hours corresponds to 3 years of observations with such a set-up.}
	 \label{fig:detcprob}
	 \end{figure*}
	 \clearpage
	 \begin{figure*}[ht] 
	  \begin{center}%
	    \includegraphics[width=15cm]{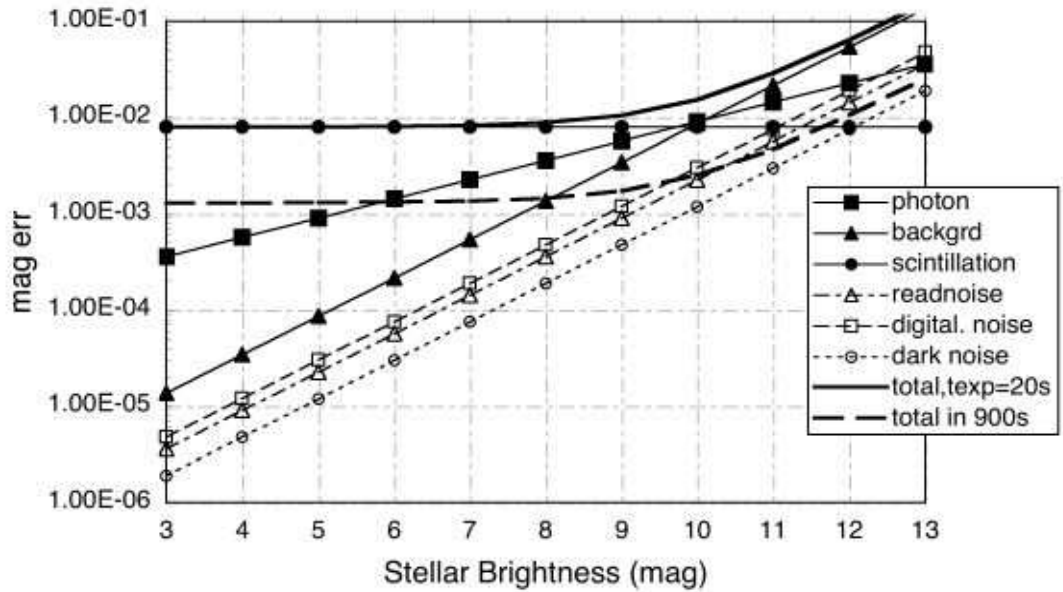} 
	     \end{center}
	     \caption{Noises of the PASS instrument against stellar magnitude, as
	     expected in star-trail images. The thick dashed line is the total noise
	     for 900 s integration, whereas all other lines indicate
	     noises in a single 20 s exposure of a field at the celestial
	     equator with the baseline set-up (see text). Photometric precision
	     suitable for transit detection can be expected up to 10.5--11 mag. Stars brighter than 5.2 mag will be saturated.}
	     \label{fig:sn}
	     \end{figure*}
	     \clearpage
	     \begin{figure*}[ht]
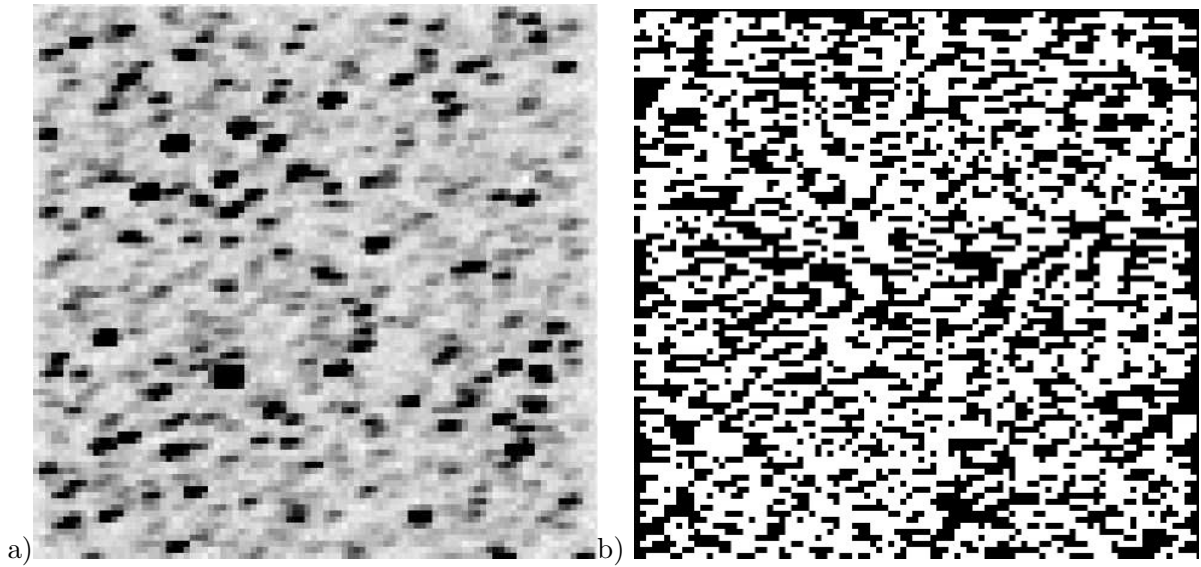
 
	      \begin{center}
	      a)%
	       \includegraphics[width=7.5cm]{figsimRV_arxiv.epsf}b) %
	       \includegraphics[width=7.5cm]{figmasks_arxiv.epsf}   
	       \end{center}
	       \caption{ a) Simulated PASS image (with stellar density typical
	       for 10$\degr$ galactic latitude), with an exposure of 20 s 
	       and the
	       baseline set-up. The size of the field is about 2$\degr$ $\times$ 
	       2$\degr$. 
	       The
	       brightest star has 5.7 mag, several have 6--9 mag, and the faintest 
	       ones
	       are 14--15 mag.  b) Final aperture mask. Starting from the
	       brightest stars, the maximum number of non-overlapping traces has 
	       been
	       fitted in (here extracting apertures for 1155 stars), so that each
	       aperture pixel (white) is assigned to just one star.}
	       \label{fig:simfield}
	       \end{figure*}
	       \clearpage
	       \begin{figure*}[ht]  
	       \begin{center}%
		\includegraphics[width=15cm]{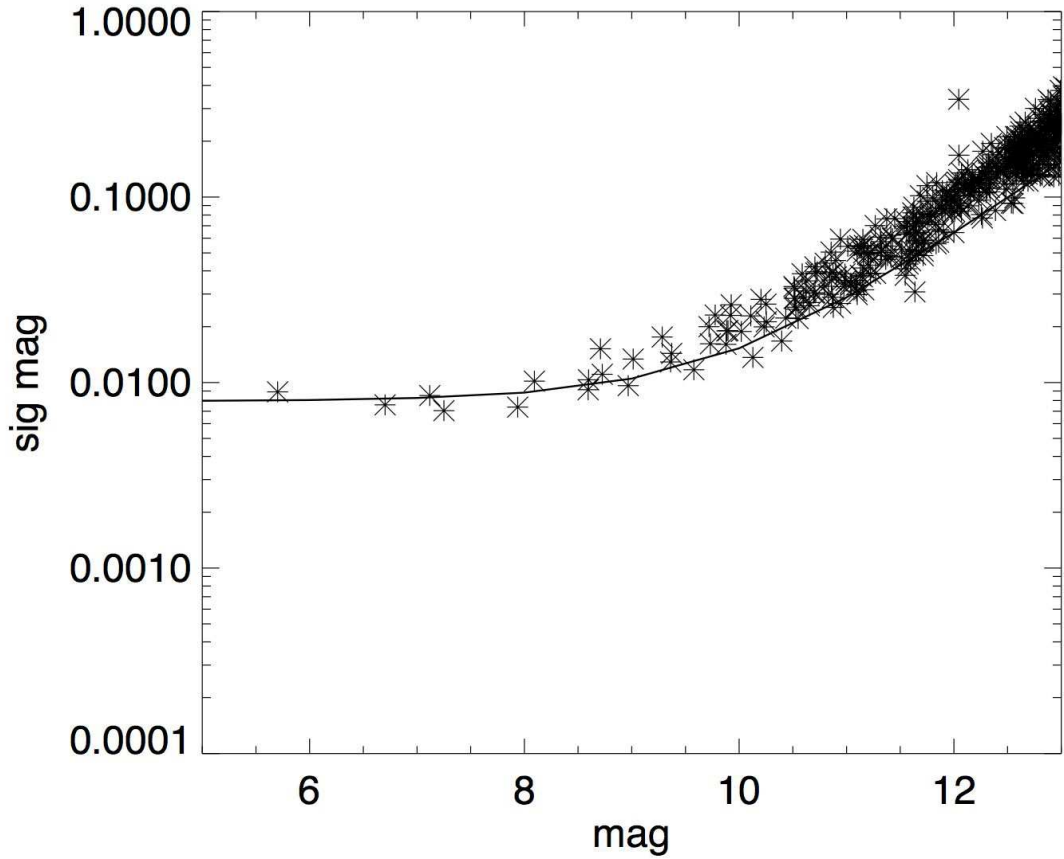} 
		 \end{center}
		 \caption{Crosses indicate the rms error found in time-series 
		 photometry
		 on the artificial stars in 15 simulated images similar to those 
		 shown
		 in Figure~\ref{fig:simfield}a. The solid line is identical to 
		 the uppermost solid line in Fig.~\ref{fig:sn}
		 and indicates the total expected noise from S/N calculations. 
		 Some
		 stars have photometric errors ``better'' than the theoretical 
		 value;
		 this is a result of the small sample of only 15 images.}
		 \label{fig:magvsmerr}
		 \end{figure*}
		 \clearpage

\begin{figure*}[ht] 
 \begin{center}%
	    \includegraphics[width=15cm]{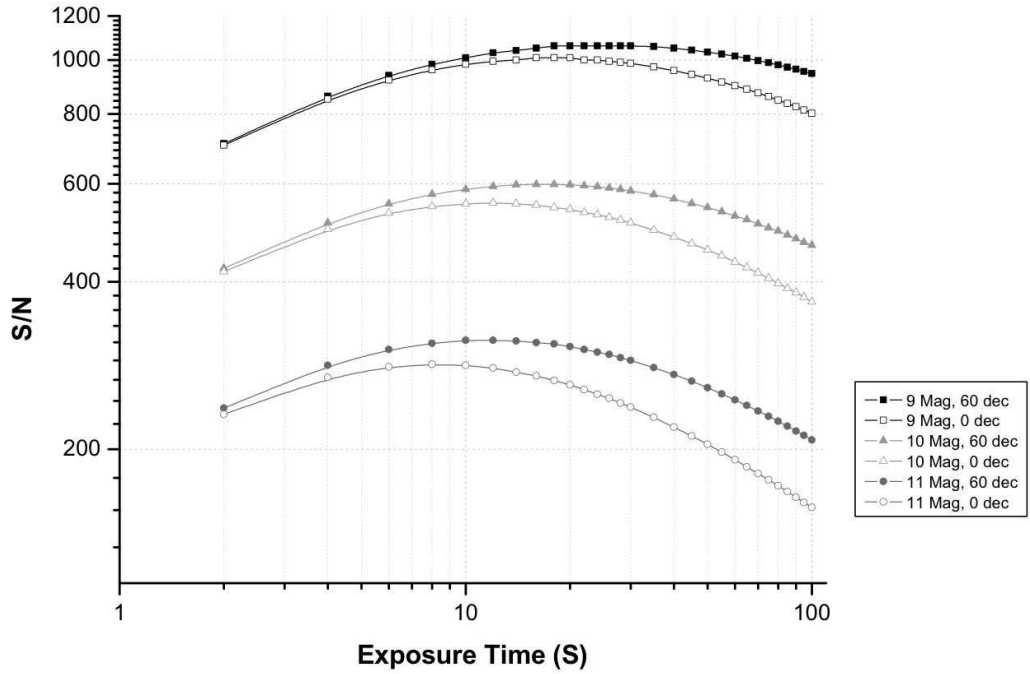} 
 \end{center}
 \caption{Dependency of the signal-to-noise of a long integration on the exposure time of individual star-trail images, as given by Eq. (B7). In the graph, an integration time of 900 second is assumed, and S/N is shown  at declinations of 0$\degr$ and 60$\degr$and for stars with magnitudes of 9, 10 and 11. The difference in S/N with declination comes from the varying velocity of the stars across the detector. Only the photon noise from star and sky-background is considered here, using values from the 'PASS-baseline'}
	     \label{fig:sntrail}
	     \end{figure*}
	     \clearpage

		 \begin{deluxetable}{crrrr}
		 \tabletypesize{\scriptsize}
		 \tablecaption{Crowding in PASS field at $l = 10\degr$ and fraction 
		 of confused apertures}
		 \tablewidth{0pt}
		 \tablehead{
		 \colhead{mag} & \colhead{Nstar} & \colhead{Nassign} & 
		 \colhead{Nconfused} & \colhead{frac$_{conf}$}}
		 \startdata4.5--5.5&1&1&0&0.000\\
		 5.5--6.5&1&1&0&0.000\\
		 6.5--7.5&3&2&1&0.333\\
		 7.5--8.5&3&3&0&0.000\\
		 8.5--9.5&9&9&0&0.000\\
		 9.5--10.5&19&18&1&0.053\\
		 10.5--11.5&76&71&5&0.066\\
		 11.5--12.5&163&132&31&0.190\\
		 12.5--13.5&462&283&179&0.387\\
		 13.5--14.5&1099&334&765&0.696\\
		 14.5--15.5&2383&301&2082&0.874\\
		 \enddata\label{tab:crowd}
		 \end{deluxetable}
		 
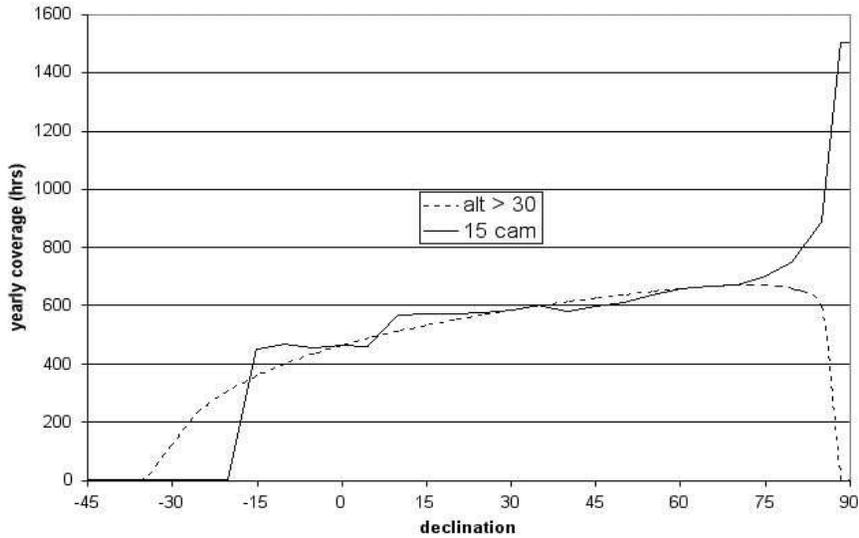
\end{document}